%% file: paper.tex
\documentclass[conference]{IEEEtran}
\usepackage{booktabs} 
\usepackage{algorithm}
\usepackage{algpseudocode}
\usepackage{graphicx}
\usepackage{caption}
\usepackage{color}
\usepackage[bookmarks=false]{hyperref}
\usepackage{amsmath}
\usepackage{amsthm}
\usepackage{amsfonts}
\usepackage{xspace}
\usepackage{multirow}
\usepackage[table,xcdraw]{xcolor}
\usepackage[center]{subfigure} 
\usepackage{bbm} 
\usepackage{tikz} 

\newcommand{\hide}[1]{}
\newcommand{\subparagraph}{} 
\usepackage{titlesec} 
\titlespacing*{\section}{0pt}{1.2pt}{1.2pt}
\titlespacing*{\subsection}{0pt}{1.2pt}{1.2pt}
\titlespacing*{\subsubsection}{0pt}{1.2pt}{1.2pt}

\begin{document}

\title{A Comparative Study of Matrix Factorization and Random Walk with Restart in Recommender Systems}

\input{author}

\maketitle

\input{dfn}

\begin{abstract}
\input{000abstract}
\end{abstract}

\begin{IEEEkeywords}
matrix factorization; random walk with restart; recommender systems
\end{IEEEkeywords}
\IEEEpeerreviewmaketitle

\section{Introduction}
\label{sec:intro}
\input{100intro}

\section{Preliminaries}
\label{sec:prelim}
\input{200prelim}

\section{Recommendation methods}
\label{sec:methods}
\input{300methods}

\section{Experiments}
\label{sec:exp}
\input{400exp}

\section{Conclusion}
\label{sec:conclusion}
\input{700conclusion}

\section*{Acknowledgment}
This work was supported by ICT R\&D program of MSIP/IITP. [2013-0-00179,
Development of Core Technology for Context-aware Deep-Symbolic Hybrid
Learning and Construction of Language Resources]. 
U Kang is the corresponding author.

\bibliographystyle{IEEEtran}
\bibliography{sigproc}

%
%

\end{document}

%% file: author.tex
\author{
	\IEEEauthorblockN{Haekyu Park}
	\IEEEauthorblockA{
		Computer Science and Engineering\\
		Seoul National University\\
		Seoul, Republic of Korea\\
		Email: hkpark627@snu.ac.kr}
	\and
	\IEEEauthorblockN{Jinhong Jung}
	\IEEEauthorblockA{
		Computer Science and Engineering\\
		Seoul National University\\
		Seoul, Republic of Korea\\
		Email: jinhongjung@snu.ac.kr}
	\and
	\IEEEauthorblockN{U Kang}
	\IEEEauthorblockA{
		Computer Science and Engineering\\
		Seoul National University\\
		Seoul, Republic of Korea\\
		Email: ukang@snu.ac.kr}
}

%% file: dfn.tex
\newtheorem{mydef}{Definition}
\definecolor{orange}{rgb}{1, 0.5, 0}
\definecolor{purple}{rgb}{0.64, 0.25, 0.84}
\definecolor{blue}{rgb}{0.3, 0.7, 1}
\floatname{algorithm}{Algorithm}
\renewcommand{\algorithmicrequire}{\textbf{Input:}}
\renewcommand{\algorithmicensure}{\textbf{Output:}}
\newtheorem{problem}{Problem}
\newcommand{\method}{UniWalk\xspace}
\newcommand{\modelname}{UniWalk}

\newcommand\blue[1]{\textcolor{blue}{#1}}
\newcommand\red[1]{\textcolor{red}{#1}}
\newcommand\orange[1]{\textcolor{orange}{#1}}
\newcommand\purple[1]{\textcolor{purple}{#1}}
\newcommand*\circled[1]{\tikz[baseline=(char.base)]{
		\node[shape=circle,draw,inner sep=0.3pt] (char) {#1};}}

\newtheorem{theorem}{Theorem}[section]
\newtheorem{corollary}{Corollary}[theorem]
\newtheorem{lemma}[theorem]{Lemma}
\renewcommand{\algorithmicrequire} {\textbf{Input:}\xspace}
\renewcommand{\algorithmicensure} {\textbf{Output:}\xspace}

\newcommand{\set}[1]{\mathbf{#1}}
\newcommand{\mat}[1]{\mathbf{#1}}
\newcommand{\matt}[1]{\mathbf{#1}^{\top}}
\newcommand{\mattt}[1]{\mathbf{\tilde{#1}}^{\top}}
\newcommand{\mati}[1]{\mathbf{#1}^{-1}}
\newcommand{\vect}[1]{\mathbf{#1}}
\newcommand{\vectt}[1]{\mathbf{#1}^{\top}}

\renewcommand{\r}[0]{\vect{r}}
\newcommand{\rt}[1]{\vect{r}^{(#1)}}
\newcommand{\q}[0]{\vect{q}}
\newcommand{\nAT}[0]{\mattt{A}}
\newcommand{\nA}[0]{\mat{\tilde{A}}}
\newcommand{\A}[0]{\mat{A}}

\newcommand{\mfe}{$\textnormal{MF}_\textnormal{Exp}$\xspace}
\newcommand{\mfi}{$\textnormal{MF}_\textnormal{Imp}$\xspace}
\newcommand{\bmf}{$\textnormal{MF}_\textnormal{Bias}$\xspace}
\newcommand{\coldmf}{$\textnormal{MF}_\textnormal{Side}$\xspace}
\newcommand{\rwr}{$\textnormal{RWR}_\textnormal{Exp}$\xspace}
\newcommand{\rwri}{$\textnormal{RWR}_\textnormal{Imp}$\xspace}
\newcommand{\rwrb}{$\textnormal{RWR}_\textnormal{Bias}$\xspace}
\newcommand{\coldrwr}{$\textnormal{RWR}_\textnormal{Side}$\xspace}

%% file: 000abstract.tex
Between matrix factorization or Random Walk with Restart (RWR), which method works better for recommender systems?
Which method handles explicit or implicit feedback data better? 
Does additional information help recommendation? 
Recommender systems play an important role in many e-commerce services such as Amazon and Netflix to recommend new items to a user.
Among various recommendation strategies, collaborative filtering has shown good performance by using rating patterns of users.
Matrix factorization and random walk with restart are the most representative collaborative filtering methods.
However, it is still unclear which method provides better recommendation performance despite their extensive utility.

In this paper, we provide a comparative study of matrix factorization and RWR in recommender systems.
We exactly formulate each correspondence of the two methods according to various tasks in recommendation.
Especially, we newly devise an RWR method using global bias term which corresponds to a matrix factorization method using biases.
We describe details of the two methods in various aspects of recommendation quality such as how those methods handle cold-start problem which typically happens in collaborative filtering.
We extensively perform experiments over real-world datasets to evaluate the performance of each method in terms of various measures.
We observe that matrix factorization performs better with explicit feedback ratings while RWR is better with implicit ones.
We also observe that exploiting global popularities of items is advantageous in the performance and that side information produces positive synergy with explicit feedback but gives negative effects with implicit one.

%% file: 100intro.tex
Recommending new items to a user has been widely recognized as an important research topic in data mining area~\cite{resnick1997recommender,ShinK14,DBLP:journals/tkde/ShinSK17,schafer1999recommender}, and recommender systems have been extensively applied in various applications across different domains to recommend items such as book~\cite{chen2005link}, movie~\cite{koren2009matrix,koren2009bellkor}, music~\cite{baltrunas2011matrix,song2012survey}, friend~\cite{naruchitparames2011friend,liben2007link}, scientific article~\cite{wang2011collaborative}, video~\cite{covington2016deep}, and restaurant~\cite{li2016point}.
Many e-commerce services such as Amazon and Netflix mainly depend on recommender systems to increase their profits by selling what consumers are interested in against overloaded information of products~\cite{linden2003amazon,shang2008personal,koren2009bellkor}.

Collaborative filtering (CF) has been successfully adopted among diverse recommendation strategies due to its high quality performance~\cite{su2009survey} and domain free property.
For a query user, CF recommends items preferred by other users who present similar rating patterns to the query user~\cite{koren2009matrix}.
This strategy does not require domain knowledge for recommendation, since it relies on only user history such as item ratings or previous transactions.
However, the domain free property causes cold-start problem:
the systems cannot recommend an item to a user if the item or the user are newly added to the system and existing data on them are not observed.
Many techniques have been proposed~\cite{Barjasteh:2015:CIU:2792838.2800196, zhou2011functional} to solve the cold-start problem,
but the problem is still a major challenge in recommender systems.

The two major approaches of collaborative filtering are \textit{latent factor models} and \textit{graph based models}.
Matrix factorization (MF)~\cite{koren2009matrix} is the most widely used method as a latent factor model, and it discovers latent factors inherent in relations between users and items by decomposing a user-item rating matrix.
Random walk with restart (RWR)~\cite{tong2006fast,KangTS12,AxiomSimTr,ShinJSK15,DBLP:journals/tods/JungSSK16,DBLP:conf/sigmod/JungPSK17} is commonly used as a graph based model for recommender systems~\cite{konstas2009social,liben2007link,DBLP:conf/icdm/JungJSK16}.
RWR recommends items to a user with the user's personalized ordering of node-to-node proximities which are measured by a random surfer in a user-item bipartite graph as shown in Figure~\ref{fig:example:rwr}.
These methods also provide solutions for the cold start problem, using side information such as user demographic information or item category data~\cite{Barjasteh:2015:CIU:2792838.2800196,zhou2011functional,konstas2009social,Yildirim:2008:RWM:1454008.1454031}.
Although MF and RWR have been extensively used with the same purpose, it is still ambiguous to answer the following question: which method is better between matrix factorization and random walk with restart in recommender systems?

This paper aims to compare matrix factorization and random walk with restart in various tasks of recommendations with the corresponding metrics.
We are interested in answering the following questions:

\begin{itemize}
	\item \textbf{Explicit feedback.} Which method performs better when explicit feedback data are given?
	\item \textbf{Implicit feedback.} Which method performs better when implicit feedback data are given?
	\item \textbf{Bias terms.} Do the bias terms improve the performance of recommendation methods? Which method performs better with bias factors?
	\item \textbf{Side information.} Does the side information improve the performance of recommendation methods? Which method is better when additional information are given?
\end{itemize}

In our experiments, matrix factorization performs better with explicit feedback data, while random walk with restart is better with implicit ones.
We also observe that biases enhance the overall quality of recommendations.
However, side information enhances the performance when used with explicit ratings, while degrades the performance with implicit ratings.
Detailed explanations are stated in Section \ref{sec:exp}.
The main contributions of this paper are as follows:
\begin{itemize}
	\item \textbf{Formulation.}
    We show that each method in matrix factorization has its corresponding one in random walk with restart. This lays the groundwork for systematic comparison of the two methods.
    \item \textbf{Method.}
    We newly devise a random walk with restart method that introduces global bias terms.
	It reflects global popularities of items and general properties of users, improving random walk with restart methods without the bias terms.
	\item \textbf{Analysis.}
	We systemically compare the two collaborative filtering approaches.
	We also analyze properties of the methods that cause their strengths in various tasks.
	\item \textbf{Experiments.} We present and discuss extensive experimental results for many scenarios with various types of input over different recommendation purposes.
\end{itemize}

The code and datasets used in this paper are available at \url{http://datalab.snu.ac.kr/mfrwr}.
The rest of this paper is organized as follows.
Section \ref{sec:prelim} explains preliminaries on matrix factorization and random walk with restart for recommender systems.
Section \ref{sec:methods} presents recommendation methods that we discuss in this paper.
Section \ref{sec:exp} shows experimental results and discussions on recommendation performance of the methods.
Finally, we conclude this paper in Section \ref{sec:conclusion}.

%% file: 200prelim.tex
Section \ref{sec:prelim} introduces preliminaries on matrix factorization and random walk with restart in recommender systems.
Table~\ref{table:notation} lists symbols and their definition used in this paper.
We denote matrices and sets with upper-case bold letters (e.g. $\mathbf{R}\text{ or }\set{U}$), vectors with lower-case bold letters (e.g. $\mathbf{x}_u$), scalars with lower-case italic letters (e.g. $c$), and graphs by upper-case normal letters (e.g. $G$).

\renewcommand{\tablename}{\textbf{Table}}
\begin{table}[h]
	\centering
	\caption{Table of symbols.}
	\label{table:notation}
	\begin{tabular}{c l}  \toprule
		\textbf{Symbol}		&	\textbf{Definition}		\\	\hline
		$\set{U}$				 &	set of users\\
		$\set{I}$				   &	set of items\\
		$\mathbf{R} \in \mathbb{R}^{|\set{U}| \times |\set{I}|}$	&	observed rating matrix	 \\
		$u$						&	user 							\\
		$i$						&	item							\\
		$r_{ui}$			&	observed rating of $u$ on $i$	\\
		$\hat{r}_{ui}$			&	predicted rating of $u$ on $i$	\\
		$\alpha$		& coefficient of confidence level in implicit feedback\\
		$\mu$				&	average of ratings	\\
		$b_u$				& bias of $u$ \\
		$b_i$				& bias of $i$ \\
		$s$						& a user attribute \\
		$a_{us}$			& user attribute of $u$ with respect to $s$\\
		$t$						& an item attribute \\
		$b_{ti}$						& item attribute of $i$ with respect to $t$ \\
		$\set{\Omega_{R}}$ & set of $(u, i)$ where $r_{ui}$ is observed \\
		$\set{\Omega_{A}}$ & set of $(u, s)$ where $a_{us}$ is observed \\
		$\set{\Omega_{B}}$ & set of $(t, i)$ where $b_{ti}$ is observed \\
		$d$            		&	dimension of latent vectors of users and items	\\
		$\mathbf{x}_u \in \mathbb{R}^d$	& vector of user $u$	\\
		$\mathbf{y}_i \in \mathbb{R}^d$	& vector of item $i$	\\
		$\mathbf{w}_s \in \mathbb{R}^d$	& vector of user attribute $s$\\
		$\mathbf{z}_t \in \mathbb{R}^d$	& vector of item attribute $t$\\
		$\lambda$			&	regularization parameter\\
		$\eta$			&	learning rate			\\
		$G=(\set{V}, \set{E})$     		& 	user-item bipartite graph\\
		$\set{V}$				& set of nodes in $G$, i.e., $\set{V}=\set{U}\cup \set{I}$\\
		$\set{E}$				& set of weighted edges $(u, i, r_{ui})$ in $G$\\
		$\textbf{A}$			& adjacency matrix of $G$ \\
		$\tilde{\textbf{A}}$	 & row-normalized adjacency matrix of $\textbf{A}$ \\
		$G'$     					& augmented graph adding side info. into $G$\\
		$\delta$				& weight of link representing side info. in $G'$ \\
		$c$							& restart probability in RWR \\
		$\textbf{q}$			& starting vector in RWR \\
		$\textbf{b}$			& bias vector in RWR\\
		$\beta$					& walk coefficient in biased RWR\\
		$\gamma$			& jump coefficient in biased RWR\\
		\bottomrule
	\end{tabular}
\end{table}

\subsection{Matrix Factorization}
\label{sec:mf}
Matrix factorization (MF) predicts unobserved ratings given observed ratings.
MF predicts a rating of item $i$ given by user $u$ as
$\hat{r}_{ui} = \mathbf{x}_u^T \mathbf{y}_i$,
where $\mathbf{x}_u$ is $u$'s vector and $\mathbf{y}_i$ is $i$'s vector.

Objective function is defined in Equation \eqref{eq:L}, where $r_{ui}$ is an observed rating and $\set{\Omega_{R}}$ is a set of (user, item) pairs
for which ratings are observed.
The term $\lambda(||\mathbf{x}_u||^2 + ||\mathbf{y}_i||^2)$ prevents overfitting by regularizing the magnitude of parameters,
where the degree of regularization is controlled by the hyperparameter $\lambda$.

\begin{align}
\hspace{-4mm}
\label{eq:L}
L =
\frac{1}{2}\sum_{(u,i) \in \set{\Omega_{R}}}
\bigg((r_{ui} - \mathbf{x}_u^T \mathbf{y}_i)^2
+ \lambda(||\mathbf{x}_u||^2 + ||\mathbf{y}_i||^2)\bigg)
\end{align}

The standard approach to learn parameters which minimize $L$ is
GD (Gradient Descent).
The update procedures for parameters in GD are as follows.

\begin{equation}
\mathbf{x}_u \gets \mathbf{x}_u - \eta{\nabla_{\mathbf{x}_u} L}, \;\;\;\;
{\nabla_{\mathbf{x}_u} L} = -e_{ui}\mathbf{y}_i + \lambda\mathbf{x}_u
\nonumber
\end{equation}

\begin{equation}
\mathbf{y}_i \gets \mathbf{y}_i - \eta{\nabla_{\mathbf{y}_i} L}, \;\;\;\;
{\nabla_{\mathbf{y}_i} L} = -e_{ui}\mathbf{x}_u + \lambda\mathbf{y}_i
\nonumber
\end{equation}
, where $\eta$ is a learning rate and $e_{ui}$ is a prediction error defined as $e_{ui} = r_{ui} - \mathbf{x}_u^T \mathbf{y}_i$.

\begin{figure}[!t]
	\centering
	\hspace{-2mm}
	\includegraphics[width=0.4\linewidth]{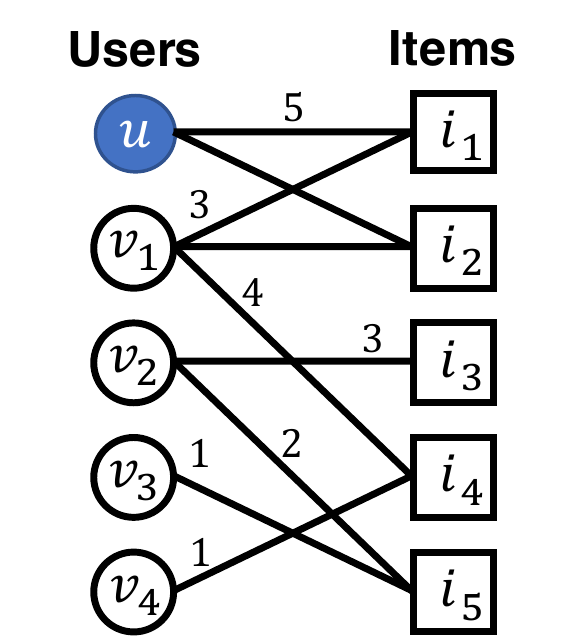}
	\caption{
		\label{fig:example:rwr}
		Example of a user-item bipartite graph. In the graph, edges between users and items are weighted with ratings.
		Node $u$ is a starting node for RWR which measures node-to-node proximities w.r.t. node $u$ to rank items.
	}
\end{figure}

\subsection{Random Walk with Restart}
\label{sec:rwr}
Random Walk with Restart (RWR) is one of the most commonly used methods for graph based collaborative filtering in recommender systems~\cite{konstas2009social,Yildirim:2008:RWM:1454008.1454031}.
Given a user-item bipartite graph $G$ and a query user $u$ as seen in Figure~\ref{fig:example:rwr}, RWR computes a personalized ranking of items w.r.t the user $u$.
The input graph $G$ comprises the set of nodes $\set{V}$ with the set of users $\set{U}$ and the set of items $\set{I}$, i.e., $\set{V}=\set{U} \cup \set{I}$.
Each edge $(u, i, r_{ui}) \in \set{E}$ represents the rating $r_{ui}$ between user $u$ and item $i$, and the rating is the weight of the edge.

RWR exploits a random surfer to produce the personalized ranking of items for a user $u$ by letting her move around the graph $G$.
Suppose the random surfer started from the user node $u$ and she is at the node $v$ currently.
Then the surfer takes one of the following actions: \emph{random walk} or \emph{restart}.
\emph{Random walk} indicates that the surfer moves to a neighbor node of the current node with probability $1-c$, and \emph{restart} indicates that the surfer goes back to the starting node $u$ with probability $c$.
If the random surfer visits node $v$ many times, then node $v$ is highly related to node $u$; thus, node $v$ is ranked high in the personalized ranking for $u$.

The random surfer is likely to frequently visit items that are rated highly by users who give ratings similarly to the query user $u$, which is consistent with the intuition of collaborative filtering, i.e., if a user has similar taste with the query user $u$, and the similar user likes an item $i$, then $u$ is likely prefer the item $i$.
We measure probabilities that the random surfer visits each item as ranking scores, called RWR scores, and sort the scores in the descending order to recommend items for the query user $u$.
The detailed method for computing RWR is described in Section~\ref{sec:method:rwr}.

%% file: 300methods.tex
\begin{table}[t]
	\centering
	\caption{
		{Correspondence between matrix factorization (MF) and random walk with restart (RWR) according to tasks.}
	}
	\begin{tabular}{l c c }
		\toprule
		Task & MF & RWR \\
		\midrule
		Explicit feedback & \mfe (Section~\ref{sec:method:mf:explicit}) & \rwr (Section~\ref{sec:method:rwr:explicit})\\
		Implicit feedback & \mfi (Section~\ref{sec:method:mf:implicit})& \rwri (Section~\ref{sec:method:rwr:implicit})\\
		Global bias       & \bmf (Section~\ref{sec:method:mf:bias})& \rwrb (Section~\ref{sec:method:rwr:bias})\\
		Using side info.        & \coldmf (Section~\ref{sec:method:mf:cold})& \coldrwr(Section~\ref{sec:method:rwr:cold})\\
		\bottomrule
	\end{tabular}
	\captionsetup{width=0.5\textwidth}
	\label{table:task}
\end{table}

Section \ref{sec:methods} describes recommendation methods to be compared in this paper.
They are based on matrix factorization (Section \ref{sec:mf}) or random walk with restart (Section \ref{sec:rwr}).
We present four methods for each approach in cases of the following scenario list.
We suggest a matrix factorization method and a random walk with restart method for each scenario, and they are summarized in Table \ref{table:task}.
\begin{itemize}
	\item  \textbf{Explicit feedback}.
	We recommend items when explicit feedback ratings are given;
	for example, movie rating data with 1 to 5 scaled ``stars" are given.
	\item \textbf{Implicit feedback}.
	We also present recommendation methods for implicit feedback data such as the number of clicks on items.
	\item \textbf{Global bias terms}.
	Bias terms indicating global properties of users and items are used in recommendation methods to predict preferences of users more accurately.
	\item \textbf{Employing side information}.
	We present recommendation methods that use auxiliary information.
	Auxiliary information of users and items is used to solve cold start problem or enhance accuracy of recommendations.
\end{itemize}

\subsection{Recommendation Methods Based on Matrix Factorization}
We explain four methods based on matrix factorization as follows.
Each of them is different in its purpose and type of datasets they use.

\subsubsection{\mfe, A Basic Matrix Factorization Method for Explicit Rating}
\label{sec:method:mf:explicit}
\mfe is a standard matrix factorization based method for explicit feedback ratings \cite{koren2009matrix}.
It predicts unobserved ratings by learning embeddings of users and items.
\mfe is described in section \ref{sec:mf}.

\subsubsection{\mfi, A Basic Matrix Factorization Method for Implicit Rating}
\label{sec:method:mf:implicit}
\mfi is for implicit feedback ratings \cite{hu2008collaborative}.
For a user $u$, an item $i$, and an implicit feedback rating $r_{ui}$,
\mfi predicts a binarized implicit feedback rating $p_{ui}$ by learning $u$'s vector $\mathbf{x}_u$ and $i$'s vector $\mathbf{y}_i$.
The implicit feedback rating $r_{ui}$ is the number of times that a user performs a favorable action to an item, e.g. the number of clicks of the item.
The binarized implicit feedback rating $p_{ui}$ is defined as 1 if $r_{ui} > 0$ and 0 otherwise.

\mfi requires confidence levels of each rating $r_{ui}$, since implicit feedback data are inherently noisy.
For example, not watching a movie occasionally indicates dislike for the movie or ignorance of its existence.
The confidence level of a rating crystallizes how certainly we can use the rating value.
We define the confidence level of an implicit rating for $u$ and $i$ as $c_{ui} = 1 + \alpha r_{ui}$ as stated in  \cite{hu2008collaborative}.

The objective function of \mfi is defined in Equation \eqref{eq:L_imp}.
$c_{ui}$ adjusts intensity of learning $\mathbf{x}_{u}$ and $\mathbf{y}_{i}$ in gradient descent update of the vectors as presented in Equations \eqref{eq:gdimpx} and \eqref{eq:gdimpy}, where $e_{ui}$ is a prediction error defined as $e_{ui} = p_{ui} - \mathbf{x}_u^T \mathbf{y}_i$.

\begin{align}
\label{eq:L_imp}
L =
\frac{1}{2}\sum_{(u,i) \in \set{\Omega_{R}}}
\bigg(c_{ui}(p_{ui} - \mathbf{x}_u^T \mathbf{y}_i)^2
+ \lambda(||\mathbf{x}_u||^2 + ||\mathbf{y}_i||^2)\bigg)
\end{align}

\begin{equation}
\label{eq:gdimpx}
\mathbf{x}_u \gets \mathbf{x}_u - \eta{\nabla_{\mathbf{x}_u} L}, \;\;\;\;
{\nabla_{\mathbf{x}_u}L} = -e_{ui} c_{ui} \mathbf{y}_i + \lambda\mathbf{x}_u
\end{equation}

\begin{equation}
\label{eq:gdimpy}
\mathbf{y}_i \gets \mathbf{y}_i - \eta{\nabla_{\mathbf{y}_i} L}, \;\;\;\;
{\nabla_{\mathbf{y}_i}L} = -e_{ui} c_{ui} \mathbf{x}_u + \lambda\mathbf{y}_i
\end{equation}

\subsubsection{\bmf, A Matrix Factorization Method with Global Bias Terms}
\label{sec:method:mf:bias}
\bmf introduces bias terms into \mfe to represent individual rating pattern of users and items.
For example, a user $u$'s bias term $b_u$ is learned to possess a high value if $u$ normally rates all items favorably.
An item $i$'s bias $b_i$ is learned to have a high value if it is rated highly by almost all users.

The objective function of \bmf is given in Equation \eqref{eq:bmfL}.
$\mu$ is a global average rating value and $b_e$ is a bias term of a user or an item $e$.

\begin{align}
\hspace{-4mm}
\label{eq:bmfL}
L =
\frac{1}{2}\sum_{(u,i) \in \set{\Omega_{R}}}
&\bigg((r_{ui} - \mu - b_u - b_i - \mathbf{x}_u^T \mathbf{y}_i)^2\\ \notag
&+ \lambda(b_u^2 + b_i^2 + ||\mathbf{x}_u||^2 + ||\mathbf{y}_i||^2)\bigg)
\end{align}

We use GD (Gradient Descent) method to minimize $L$ in Equation \eqref{eq:bmfL}.
The update procedures for parameters are as follows.

\begin{equation}
b_u \gets b_u - \eta{\partial L \over \partial b_u}, \;\;\;\;
{\partial L \over \partial b_u} = -e_{ui} + \lambda b_u
\nonumber
\end{equation}
\begin{equation}
b_i \gets b_i - \eta{\partial L \over \partial b_i}, \;\;\;\;
{\partial L \over \partial b_i} = -e_{ui} + \lambda b_i
\nonumber
\end{equation}
\begin{equation}
\mathbf{x}_u \gets \mathbf{x}_u - \eta{\nabla_{\mathbf{x}_u} L}, \;\;\;\;
{\nabla_{\mathbf{x}_u}L} = -e_{ui}\mathbf{y}_i + \lambda\mathbf{x}_u
\nonumber
\end{equation}

\begin{equation}
\mathbf{y}_i \gets \mathbf{y}_i - \eta{\nabla_{\mathbf{y}_i} L}, \;\;\;\;
{\nabla_{\mathbf{y}_i}L} = -e_{ui}\mathbf{x}_u + \lambda\mathbf{y}_i
\nonumber
\end{equation}
, where $e_{ui}$ is a prediction error defined as $e_{ui} = r_{ui} - \mu - b_u - b_i - \mathbf{x}_u^T \mathbf{y}_i$.

\subsubsection{\coldmf, A Coupled Matrix Factorization Method Using Side Information}
\label{sec:method:mf:cold}
\coldmf uses additional information to understand users and items in many-sided properties.
User similarity information such as friendship in a social network or item similarity information such as items' category is advantageous for finding users with similar tastes and items with similar properties, which causes latent vectors of users and items to be similar.
Side information is useful when rating data of users and items are missing.
For example, if a cold-start user $u$ has similar demographic information with another warm-user $v$,
$\mathbf{x}_u$ is able to be learned to be similar to $\mathbf{x}_v$ even though $u$ has never rated an item.

The objective function of \coldmf is defined in Equation \eqref{eq:coldL}.
\vspace{-1em}
\begin{align}
\label{eq:coldL}
L=
&
\overbrace{
	\frac{1}{2}\sum_{(u,i) \in \set{\Omega_R}}
	\bigg((r_{ui} - \mathbf{x}_u^T \mathbf{y}_i)^2
    + \lambda(||\mathbf{x}_u||^2 + ||\mathbf{y}_i||^2)\bigg)
}
^{\circled{1} \text{ Factorization of ratings}}
\notag\\
+&
\overbrace{
	\frac{1}{2}\sum_{(u,s) \in \set{\Omega_A}}
	\bigg((a_{us} - \mathbf{x}_u^T \mathbf{w}_s)^2
	+ \lambda(||\mathbf{x}_u||^2 + ||\mathbf{w}_s||^2)\bigg)
}
^{\circled{2} \text{ Factorization of user similarity attributes}}
\notag\\
+&
\overbrace{
	 \frac{1}{2}\sum_{(t,i) \in \set{\Omega_B}}
	\bigg((b_{ti} - \mathbf{z}_t^T \mathbf{y}_i)^2
	+ \lambda(||\mathbf{z}_t||^2 + ||\mathbf{y}_i||^2)\bigg)
}
^{\circled{3} \text{ Factorization of item similarity attributes}}
\end{align}

$\set{\Omega_R}$ is a set of (user, item) where ratings are observed.
$\set{\Omega_A}$ is a set of (user, user attribute) where the user similarity attributes are observed.
$\set{\Omega_B}$ is a set of (item attribute, item) where the item similarity attributes are observed.
For a user $u$ and a user similarity attribute $s$, $a_{us}$ indicates $u$'s properties on $s$ such as $u$'s age.
$\mathbf{w}_s$ is a latent vector of $s$.
$\mathbf{x}_u$ is $u$'s latent feature that couples ratings and user attributes.
For an item $i$ and an item similarity attribute $t$, $b_{ti}$ is $i$'s attribute value with respect to $t$ such as $i$'s category if $t$ indicates category of items.
$\mathbf{z}_t$ is a latent vector of $t$, and $\mathbf{y}_i$ is a latent vector of $i$ that represents ratings and item similarity attributes  of $i$ simultaneously.

We use the following gradient descent updates for $\mathbf{x}_u$, $\mathbf{y}_i$, $\mathbf{w}_s$, and $\mathbf{z}_t$.
For a user $u$ and an item $i$ for $(u, i) \in \set{\Omega_R}$, we update $\mathbf{x}_u$ and $\mathbf{y}_i$ as follows, where $e_{ui} = r_{ui} - \mathbf{x}_u^T\mathbf{y}_i$.
\begin{equation}
\mathbf{x}_u \gets \mathbf{x}_u - \eta{\nabla_{\mathbf{x}_u} \circled{1}}, \;\;\;\;
{\nabla_{\mathbf{x}_u} \circled{1}} = -e_{ui}\mathbf{y}_i + \lambda\mathbf{x}_u
\nonumber
\end{equation}
\begin{equation}
\mathbf{y}_i \gets \mathbf{y}_i - \eta{\nabla_{\mathbf{y}_i} \circled{1}}, \;\;\;\;
{\nabla_{\mathbf{y}_i} \circled{1}} = -e_{ui}\mathbf{x}_u + \lambda\mathbf{y}_i
\nonumber
\end{equation}
For a user $u$ and a user similarity attribute $s$ for $(u, s) \in \set{\Omega_A}$, we update $\mathbf{x}_u$ and $\mathbf{w}_s$ as follows, where $e_{us} = a_{us} - \mathbf{x}_u^T\mathbf{w}_s$.
\begin{equation}
\mathbf{x}_u \gets \mathbf{x}_u - \eta{\nabla_{\mathbf{x}_u} \circled{2}}, \;\;\;\;
{\nabla_{\mathbf{x}_u} \circled{2}} = -e_{us}\mathbf{w}_s + \lambda\mathbf{x}_u
\nonumber
\end{equation}
\begin{equation}
\mathbf{w}_s \gets \mathbf{w}_s - \eta{\nabla_{\mathbf{w}_s} \circled{2}}, \;\;\;\;
{\nabla_{\mathbf{w}_s} \circled{2}} = -e_{us}\mathbf{x}_u + \lambda\mathbf{w}_s
\nonumber
\end{equation}
For an item similarity attribute $t$ and an item $i$ for $(t, i) \in \set{\Omega_B}$, we update $\mathbf{z}_t$ and $\mathbf{y}_i$ as follows, where $e_{ti} = b_{ti} - \mathbf{z}_t^T\mathbf{y}_i$.
\begin{equation}
\mathbf{z}_t \gets \mathbf{z}_t - \eta{\nabla_{\mathbf{z}_t} \circled{3}}, \;\;\;\;
{\nabla_{\mathbf{z}_t} \circled{3}} = -e_{ti}\mathbf{y}_i + \lambda\mathbf{z}_t
\nonumber
\end{equation}
\begin{equation}
\mathbf{y}_i \gets \mathbf{y}_i - \eta{\nabla_{\mathbf{y}_i} \circled{3}}, \;\;\;\;
{\nabla_{\mathbf{y}_i} \circled{3}} = -e_{ti}\mathbf{z}_t + \lambda\mathbf{y}_i
\nonumber
\end{equation}

\subsection{Recommendation Methods Based on Random Walk with Restart}
\label{sec:method:rwr}
We explain four methods based on random walk with restart according to rating types of input data and their purposes.

\subsubsection{\rwr, A Basic Random Walk with Restart Method for Explicit Rating}
\label{sec:method:rwr:explicit}
Suppose we have a user-item bipartite graph $G=(\set{V}, \set{E})$ where $\set{V}$ is the set of nodes, and $\set{E}$ is the set of edges.
$\set{V}$ consists of users and items, i.e., $\set{V} = \set{U} \cup \set{I}$ where $\set{U}$ is the set of users and $\set{I}$ is the set of items.
Each edge $(u, i, r_{ui}) \in \set{E}$ represents a rating between user $u$ and item $i$, and the edge is weighted with the rating $r_{ui}$.
For a starting node $u$, the RWR scores are defined as the following recursive equation~\cite{tong2006fast}:
\begin{equation}
	\r = (1-c)\nAT\r + c \q
	\label{eq:rwr}
\end{equation}
\noindent where $\r \in \mathbb{R}^{|\set{V}|}$ is the RWR score vector w.r.t. the starting node $u$, $\q \in \mathbb{R}^{|\set{V}|}$ is the starting vector whose $u$-th entry is $1$ and all other entries are 0, and $c$ is the restarting probability.
$\nA$ is the row-normalized adjacency matrix, i.e., $\nA = \mati{D}\A$ where $\A$ is the weighted adjacency matrix of the graph $G$, and $\mat{D}$ is a diagonal matrix of weighted degrees such that $\mat{D}_{ii} = \sum_{j}\A_{ij}$ and $\A_{ij} = r_{ij}$ indicating the weighted edge $(i,j, r_{ij}) \in \set{E}$.

The RWR score vector $\r$ is iteratively updated as follows:
\begin{equation}
	\label{eq:rwr:update}
	\rt{t} \leftarrow (1-c)\nAT\rt{t} + c\q
\end{equation}
\noindent where $\rt{t}$ is the RWR score vector of $t$-th iteration.
The iteration starts with the initial RWR score vector $\rt{0}$, and it is repeated until convergence (i.e., the iteration stops when $\lVert \rt{t} - \rt{t-1} \rVert < \epsilon$ where $\epsilon$ is the error tolerance).
The iteration for $\rt{t}$ converges to a unique solution~\cite{langville2011google}.
We initialize $\rt{0}$ as $\frac{1}{|\set{V}|}\vect{1}$ where $|\set{V}|$ is the number of nodes and $\vect{1} \in \mathbb{R}^{|\set{V}|}$ is an all-ones vector.
Note that for each user, we compute the RWR score vector $\r$, and rank items in the order of the RWR scores on items $\r_{i}$ s.t. $i \in \set{I}$.

\begin{figure}[!t]
	\centering
	\hspace{-2mm}
	\subfigure[Cold-start problem]
	{
		\label{fig:example:rwr_cs:cold_start}
		\includegraphics[width=0.5\linewidth]{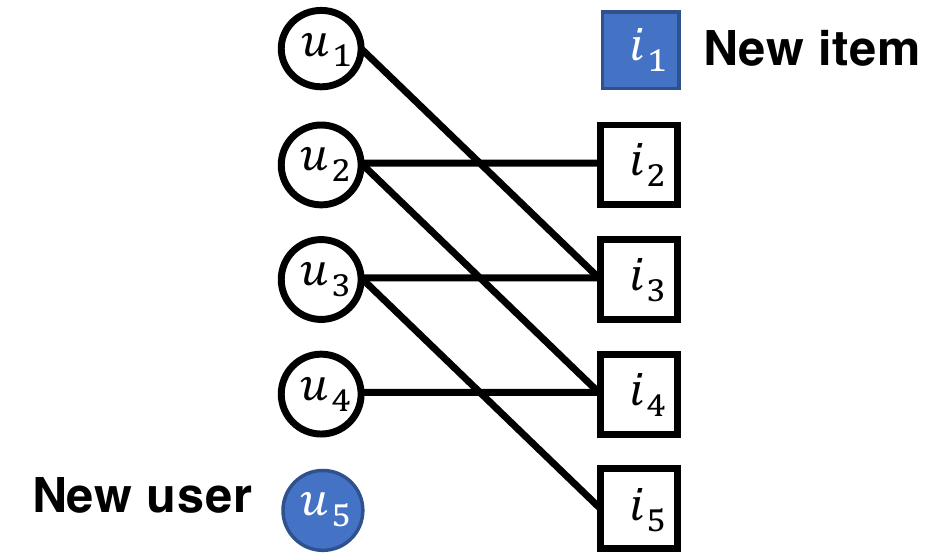}
	}
	\subfigure[Augmented graph using side information]
	{
		\label{fig:example:rwr_cs:augmented_graph}
		\includegraphics[width=0.6\linewidth]{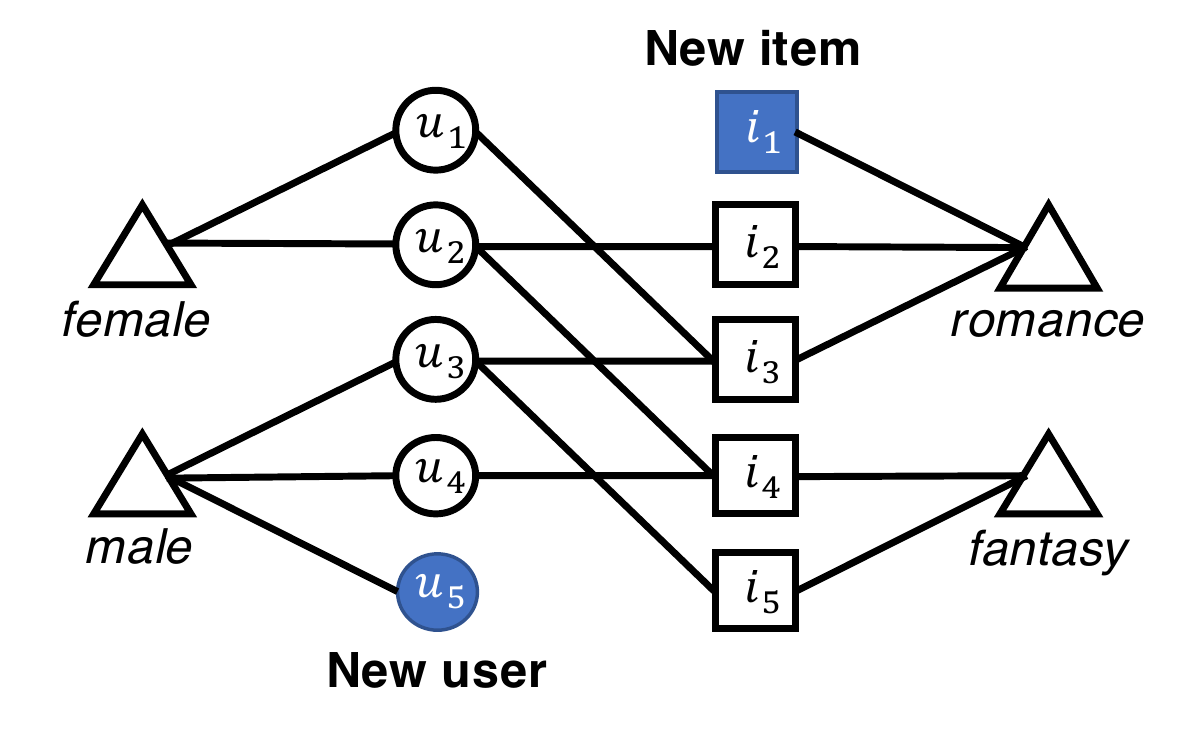}
	}
	\caption{
		\label{fig:example:rwr_cs}
		Examples of the cold-start problem in a user-item bipartite graph (a) and the augmented bipartite graph (b).
		In (a), user $u_5$ and item $i_1$ have the cold-start problem since they do not have edges with other users or items.
		(b) shows how the augmented graph resolves the cold-start problem using side information such as user demographic information or item category data.
	}
\end{figure}

\subsubsection{\rwri, A Basic Random Walk with Restart Method for Implicit Rating}
\label{sec:method:rwr:implicit}

When explicit user-item ratings are not available, RWR is also able to exploit implicit feedback ratings described in Section~\ref{sec:method:mf:implicit} for recommendation.
Instead of constructing the user-item bipartite graph using the explicit ratings, we build the graph $G$ based on the implicit feedback ratings.
An edge between user $u$ and item $i$ in the graph $G$ is represented as $(u, i, c_{ui})$ indicating an implicit rating for user $u$ and item $i$.
$c_{ui}$ is the confidence level of the implicit rating as explained in Section~\ref{sec:method:mf:implicit}.
Then we compute RWR scores in the user-item bipartite graph $G$ using Equation~\eqref{eq:rwr:update}.

\subsubsection{\rwrb, A Random Walk with Restart Method with Global Bias Terms}
\label{sec:method:rwr:bias}
As described in Section~\ref{sec:method:rwr:explicit}, the traditional RWR methods do not consider global user and item popularity even if the popularity affects the performance of recommendation as discussed in \cite{koren2009matrix}.
Thus we propose a novel random walk with restart method \rwrb for considering the global popularity (or bias) for users and items as well as personalized information in recommendation.

For a query user $u$, the goals of \rwrb are stated as follows:
(1) it obtains RWR scores of other nodes with respect to $u$ which indicate similarity with $u$,
and (2) it considers global popularities of users and items in RWR scores.
\rwrb builds on top of a basic RWR method by adding bias terms to achieve the above goals.
The first goal of calculating similarity score is done by the basic RWR approach.
Bias terms are implemented to achieve the second goal to increase RWR scores of generally popular items or users while decreasing the scores of generally unpopular ones.

The iterative equations to update RWR score vector and bias vectors are defined in Equations \eqref{eq:rwrr} and \eqref{eq:rwrb}, respectively.
The \rwrb score vector $\r$ consists of RWR terms as $\beta {\nAT\r} + \gamma \q$ and bias terms $(1 - \beta - \gamma) \vect{b}$.
The bias vector $\vect{b} \in \mathbb{R}^{|\set {V}|}$ consists of the random walk term as $(1 - c) \nAT \vect{b}$ and a jump term as $c\tilde{\vect{m}}$ where $\tilde{\vect{m}}$ is aimed to increase RWR scores of popular nodes by assigning higher probabilities that a random surfer jumps to the popular entities.
$\tilde{\vect{m}}$ is normalized to sum to 1 from a non-normalized vector $\vect{m}$; 
for each entity $e$, $e$-th entry value in $\vect{m}$ is the global average rating value of the corresponding type of entity $e$, i.e., the average of the ratings that an item receives or the average of the ratings that a user gives.

We first compute the bias vector $\vect{b}$ as follows:
\begin{equation}
	\vect{b} = (1 - c) \nAT \vect{b} + c \tilde{\vect{m}} \Leftrightarrow \vect{b} = \mat{G'}\vect{b}
	\label{eq:rwrb}
\end{equation}
\noindent where $\mat{G'} = (1-c)\nAT + c\tilde{\vect{m}}\vectt{1}$ and $\vect{1}$ is an all-ones column vector.
Note that $\mat{G'}$ is a column stochastic matrix if $\nAT$ is column stochastic, since $\vectt{1}\mat{G'}=\vectt{1}$.
Also the sum of each entry of $\vect{b}$ is 1 (i.e., $\vectt{1}\vect{b}=1$), which is proved as follows:
\begin{align}
	\label{eq:proof:sum_b}
	\begin{split}
		&\vectt{1}\vect{b} = (1-c)\vectt{1}\nAT\vect{b} + c\vectt{1}\vect{\tilde{m}} = (1-c)\vectt{1}\vect{b} + c   \\
		&\Leftrightarrow c\vectt{1}\vect{b} = c \Leftrightarrow \vectt{1}\vect{b} = 1
	\end{split}
\end{align}

Then, we compute the \rwrb score vector $\r$ as follows:
\begin{equation}
	\r = \beta {\nAT\r} + \gamma \q + (1 - \beta - \gamma) \vect{b} \Leftrightarrow \vect{r} = \mat{G}\vect{r}
	\label{eq:rwrr}
\end{equation}
\noindent where $\mat{G} = \beta\nAT + \gamma \vect{q}\vectt{1} + (1 - \beta - \gamma)\vect{b}\vectt{1}$.
Note that $\mat{G}$ is also column stochastic if $\nAT$ is a column stochastic matrix since $\vectt{1}\mat{G} = \vectt{1}$, and $\vectt{1}\vect{r} = 1$ which is proved similarly to Equation~\eqref{eq:proof:sum_b}.

\subsubsection{\coldrwr, A Random Walk with Restart Method Using Side Information}
\label{sec:method:rwr:cold}
The basic RWR described in Section~\ref{sec:method:rwr:explicit} also suffers from the cold-start problem since there are no out-going edges from a new user or a new item as shown in Figure~\ref{fig:example:rwr_cs:cold_start}.
In this case, a random surfer is stuck in the new user node or the surfer cannot reach at the new item node; thus, the basic RWR cannot compute a personalized ranking of items for the new user, and the newly added item is omitted from recommendation list.

The main idea to solve the cold-start problem is to augment the user-item bipartite graph $G$ using side information such as user demographic information or item category data~\cite{konstas2009social,Yildirim:2008:RWM:1454008.1454031}.
Figure~\ref{fig:example:rwr_cs:augmented_graph} depicts the example of how to augment the graph using side information.
Suppose we have gender information of users. Then we add \textit{female} and \textit{male} nodes into the graph $G$, and connect gender nodes and users nodes as in Figure~\ref{fig:example:rwr_cs:augmented_graph}.
Item nodes are also augmented similarly, using item information such as item category data.

Let $G'$ denote the augmented graph.
We add additional information into $G'$ as links with weights set to $\delta$.
Then \coldrwr computes RWR scores on the augmented graph $G'$ based on Equation~\eqref{eq:rwr}.
Note that in Equation~\eqref{eq:rwr}, we normalize the adjacency matrix of the augmented graph $G'$.
\coldrwr also ranks items in the order of the RWR scores on items $\r_{i}$ s.t. $i \in \set{I}$ to generate an item recommendation list for each user.

%% file: 400exp.tex
We conduct extensive experiments to compare recommendation methods which are based on matrix factorization or random walk with restart.
We focus on the following research questions according to the major tasks and present answers we got from the experiments.

\begin{itemize}
	\item \textbf{Explicit feedback} (Section \ref{sec:exp:exp})
		\vspace{-0.2em}
		\begin{itemize}
			\item Q1. \mfe vs. \rwr: Which method performs better when using explicit feedback data?
				\begin{itemize}
					\vspace{-0.2em}
					 \item A1. \mfe is better.
				\end{itemize}
		\end{itemize}
	\vspace{-0.2em}
	\item \textbf{Implicit feedback} (Section \ref{sec:exp:imp})
		\begin{itemize}
			\vspace{-0.2em}
			\item Q2. \mfi vs. \rwri: Which method performs better when using implicit feedback data?
				\begin{itemize}
					\vspace{-0.2em}
					\item A2. \rwri is better.
				\end{itemize}
		\end{itemize}
	\vspace{-0.2em}
	\item \textbf{Global bias terms} (Section \ref{sec:exp:bias})
		\begin{itemize}
			\vspace{-0.2em}
			\item Q3. Do global bias terms improve performance?
				\begin{itemize}
					\vspace{-0.2em}
					\item A3. Yes.
				\end{itemize}
			\vspace{-0.2em}
			\item Q4. \bmf vs. \rwrb: Which method performs better when exploiting global bias terms?
				\begin{itemize}
					\vspace{-0.2em}
					\item A4. \bmf is better.
				\end{itemize}
		\end{itemize}
	\vspace{-0.2em}
	\item \textbf{Employing side information} (Section \ref{sec:exp:side})
		\begin{itemize}
			\item Q5. Does side information enhance performance?
				\begin{itemize}
					\vspace{-0.2em}
					\item A5. Yes with explicit ratings, and No with implicit ratings.
				\end{itemize}
			\vspace{-0.2em}
			\item Q6. \coldmf vs. \coldrwr: Which method performs better when employing side information?
				\begin{itemize}
					\vspace{-0.2em}
					\item A6. \coldmf is better with explicit rating data and \coldrwr is better with implicit rating data.
				\end{itemize}
			\vspace{-0.2em}
			\item Q7. \coldmf vs. \coldrwr: Which method solves the cold start problem better when employing side data?
				\begin{itemize}
					\vspace{-0.2em}
					\item A7. \coldmf is better in explicit rating data and \coldrwr is better in implicit rating data.
				\end{itemize}
		\end{itemize}		
\end{itemize}

\subsection{Experimental Setup}
\subsubsection{Datasets}
We use various rating datasets such as explicit rating data, implicit rating data, and ratings with additional information.
We summarize the datasets in Tables~\ref{table:datasetstat} and~\ref{table:sidestat}.

\begin{itemize}
	\item \textit{Movielens} (\url{https://movielens.org}) is a movie recommendation site.
	It contains 0.9K users and 1.6K items.
	Ratings are in a range from 1.0 to 5.0, and their unit interval is 1.0.
	It provides user demographic information such as age, gender, occupation, and zip code.
	The dataset is available at \url{https://grouplens.org/datasets/movielens/100k/}.

	\item \textit{FilmTrust} is a movie sharing and rating website.
	This dataset is crawled by Guo et al. \cite{guo2013novel}.
	It contains 1.6K users and 2K items where users are connected by directed edges in a trust network.
	Ratings are in a range from 0.5 to 4.0, and their unit interval is 0.5.
	The dataset is available at \url{https://www.librec.net/datasets.html}.
	
	\item \textit{Epinions} (\url{epinions.com})
	is a who-trust-whom online social network of general consumer reviews.
	It contains 49K users and 140K items where users are connected by directed edges in a trust network.
	Ratings are in a range from 1.0 to 5.0, and their unit interval is 1.0.
	This dataset is available at \url{http://www.trustlet.org/epinions.html}.
	
	\item \textit{Lastfm} (\url{https://www.last.fm})
	is an online music site for free music data such as music, video, photos, concerts, and so on.
	It includes 1.9K users, 17.6K items, 92.8K implicit ratings, and 25.4K social links.
	Items are artists in this dataset, and ratings are the number of times that a user listens to the artists' music.
	The dataset is available at \url{https://grouplens.org/datasets/hetrec-2011/}.

	\item \textit{Audioscrobbler} (\url{http://www.audioscrobbler.net})
	is a database system that tracks people's habits of listening to music.
	The dataset contains 148K users, 1,631K items, 24,297K implicit ratings.
	Side information is not provided.
	The dataset is available at \url{http://www-etud.iro.umontreal.ca/~bergstrj/audioscrobbler_data.html}.
\end{itemize}

\begin{table}[t]
	\centering
	\caption{Statistics of datasets.}
	\renewcommand{\tabcolsep}{1.7mm}
	\begin{tabular}{l r r r c}
		\toprule
			& \textbf{\# of users} & \textbf{\# of items} & \textbf{\# of ratings} & \textbf{rating types} \\
		\midrule
			\textbf{Movielens} & 943 & 1,682 & 100,000 & explicit \\
			\textbf{FilmTrust} & 1,642 & 2,071 & 35,494 & explicit \\
			\textbf{Epinions} & 49,289 & 139,738 & 664,824 & explicit \\
		\midrule
			\textbf{Lastfm} & 1,892 & 17,632 & 92,834 & implicit \\
			\textbf{Audioscrobbler} & 148,111 & 1,631,028 & 24,296,858 & implicit \\
		\bottomrule
	\end{tabular}
	\label{table:datasetstat}
\end{table}

\begin{table}[t]
	\centering
	\caption{Description of side information of datasets.}
	\renewcommand{\tabcolsep}{1.2mm}
	\begin{tabular}{l l l}
		\toprule
			& \textbf{types of side info} & \textbf{details} \\
		\midrule
			\multirow{2}{*}{\textbf{Movielens}} & user demographic & age, gender, occupation, and zip code \\
							& \multicolumn{1}{l}{information}    & \multicolumn{1}{l}{of all users}     \\
			\textbf{FilmTrust} & social network  &  1,309 social links  \\
			\textbf{Epinions}   & social network  &  664,824 social links \\
			 \textbf{Lastfm}      & social network &  25,434 social links \\
			\textbf{Audioscrobbler} & N/A            & N/A \\
		\bottomrule
	\end{tabular}
	\label{table:sidestat}
\end{table}

\subsubsection{Cross-validation}
We separate training and test sets under 5-folded cross validation for all experiments except other ones related to the cold start problem; the detailed setting for the cold start problem is in Section~\ref{sec:exp:side}.
It ensures that each observed rating is included in a test set only once.

\subsubsection{Hyperparameters}

We find out hyperparameters which give the best performance for each experiment.
They are set as follows.
We use $\alpha=0.0001$ for Lastfm and Audioscrobbler in all case, and $d=5$ for all matrix factorization methods.
In \mfe, \mfi, and \bmf: $\lambda=0.3, \eta=0.05$ for all datasets.
In \coldmf: $\lambda=0.3, \eta=0.01$ for Movielens, $\lambda=0.25, \eta=0.02$ for FilmTrust,  $\lambda=0.25, \eta=0.03$ for Epinions,  and $\lambda=0.3, \eta=0.05$ for Lastfm.
In \rwr: $c=0.2$ for Movielens and FilmTrust dataset, and $c=0.5$ for Epinions dataset.
In \rwri: $c=0.1$ for all datasets.
In \rwrb: $\beta=0.4, \gamma=0.3, c=0.2$ for Movielens, $\beta=0.25, \gamma=0.1, c=0.2$ for FilmTrust, $\beta=0.5, \gamma=0.3, c=0.2$ for Epinions.
In \coldrwr:$\delta=2.0, c=0.2$ for Movielens, $\delta=1.0, c=0.2$ for Filmtrust, Epinions, and Lastfm.

\subsection{Performance Measures}
We employ two measures to compare performance of matrix factorization and random walk with restart.
We measure global ranking (Section \ref{sec:exp:ranking}) and top-$k$ prediction performance (Section \ref{sec:exp:topn}).
Global ranking and top-$k$ prediction are sufficiently important measures, since recommendations provided by e-commerce services are usually ranked list of items and customers pay attention only to the top few items.
Matrix factorization is known to optimize the lowest RMSE (Root Mean Square Error).
However, we do not use RMSE as accuracy measure because RWR cannot generate predicted ratings in the same scale with the observed data.

\subsubsection{Ranking performance}
\label{sec:exp:ranking}
We compare recommendation methods with respect to how well they predict rankings of recommended items.
We use Spearman's $\rho$ as the global ranking performance metric,
which compares a ranked list with a ground-truth ranked list and shows correlation between the lists.
$\rho$ has a value within $[-1, 1]$, and a higher value of it tells the rank of the two lists are more similar.

$\rho$ is defined as average of $\rho_u$ for all user $u$ as stated in Equation \eqref{eq:entirecorr}.
$\rho_u$ is defined in Equation \eqref{eq:corr}.
$\set{\Omega_R}^{test}[u]$ is a set of items that $u$ gives ratings in test set,
$s_{ui}$ is the rank of $i$ in a list of items sorted by predicted ratings,
$s_{ui}^*$ is the rank of $i$ in a list of items ranked by actual ratings in test set,
$\bar{s}_{ui}$ is the average of $s_{ui}$ over all items $i \in \set{\Omega_R}^{test}[u]$,
and $\bar{s}_{ui}^*$ is the average of $s_{ui}^*$ over all items $i \in \set{\Omega_R}^{test}[u]$.
If there are items that are tied for the same ranking,
we define $s_{ui}$ and $s_{ui}^*$ as the average rank of all items of the same scores with $i$.

\begin{equation}
\rho = \frac{1}{| \set{U} |} \sum_{u \in \set{U}} \rho_u
\label{eq:entirecorr}
\end{equation}
\begin{equation}
\rho_u = \frac{\sum_{i \in \set{\Omega_R}^{test}[u]} (s_{ui} - \bar{s}_{ui}) (s_{ui}^* - \bar{s}_{ui}^*)}
						{\sqrt{\sum_{i \in \set{\Omega_R}^{test}[u]} (s_{ui} - \bar{s}_{ui})^2}
						  \sqrt{\sum_{i \in \set{\Omega_R}^{test}[u]} (s_{ui}^* - \bar{s}_{ui}^*)^2} }
\label{eq:corr}
\end{equation}

\subsubsection{Top-$k$ prediction performance}
\label{sec:exp:topn}
We compare methods in performance on predicting top-$k$ recommended items.
We measure $precision@k$ where $k$ is the number of top items of interest.
It is ratio of the number of actual positive items among the first $k$ items in recommendation list predicted by methods.
The values are in range of [0, 1], and higher values indicate better predictive performance.

$precision@k$ is the average of $precision@k_u$ for all users $u$ as defined in Equation \eqref{eq:patk}.
For a user $u$, $precision@k_u$ is defined in Equation \eqref{eq:patku}.
$Actual_u(k)$ is a set of top-$k$ items sorted by observed ratings given by $u$ in test set,
and $Predicted_u(k)$ is that of top-$k$ items in test set predicted by a recommendation method.
\begin{equation}
	precision@k = \frac{1}{| \set{U} |} \sum_{u \in \set{U}} precision@k_u
	\label{eq:patk}
\end{equation}
\begin{equation}
precision@k_u = \frac{|Actual_u(k)  \cap Predicted_u(k)|}{k}
\label{eq:patku}
\end{equation}

\begin{table*}[!t]
	\centering
	\caption{
		Comparison of matrix factorization and random walk with restart in using \textbf{explicit feedback} datasets.
		\mfe shows better performance than \rwr in global ranking and top-1, 2, 3 prediction measure, as \mfe presents higher Spearman's $\rho$ and $precision@1, 2, 3$ than \rwr does.
	}
	\label{table:explicit}
	\renewcommand{\tabcolsep}{0.9mm}
	\begin{tabular}{c | c | ccc | ccc | ccc | ccc}
	\toprule
		& Performance & \multicolumn{3}{c | }{Spearman's $\rho$} & \multicolumn{3}{c | }{$precision@1$}  & \multicolumn{3}{c | }{$precision@2$}  & \multicolumn{3}{c}{$precision@3$}  \\ \cline{3-14}
		& summary     & Movielens  & FilmTrust  & Epinions & Movielens & FilmTrust & Epinions & Movielens & FilmTrust & Epinions & Movielens & FilmTrust & Epinions \\
	\midrule
		\mfe  & Higher      & \textbf{0.328}      & \textbf{0.377}      & \textbf{0.762}    & \textbf{0.144}     & \textbf{0.381}     & \textbf{0.597}    & \textbf{0.260}     & \textbf{0.514}     & \textbf{0.653}    & \textbf{0.365}     & \textbf{0.573}     & \textbf{0.624}    \\
		\rwr & Lower       & 0.238      & 0.359      & 0.616    & 0.103     & 0.319     & 0.498    & 0.197     & 0.470     & 0.585    & 0.306     & 0.549     & 0.587   \\
	\bottomrule
	\end{tabular}
\end{table*}

\subsection{Performance of Methods with Explicit Feedback Data}
\label{sec:exp:exp}
We compare performance of matrix factorization and random walk with restart when explicit feedback ratings are given.
We find out that \mfe performs better than \rwr in this situation.
Table \ref{table:explicit} presents the results of the experiment.
In using explicit feedback ratings, \mfe performs better than \rwr in global ranking and top-$k$ prediction measures.
\mfe shows higher Spearman's $\rho$ for global ranking performance and also higher $precision@1, 2, 3$ measures for top-1, 2, 3 prediction measures for all explicit rating datasets.

\subsection{Performance of Methods with Implicit Feedback Data}
\label{sec:exp:imp}
We compare matrix factorization and random walk with restart in their global ranking and top-$k$ prediction performance when implicit feedback ratings are given.
We find out that \rwri performs better than \mfi in using implicit rating data.
Table \ref{table:implicit} tells the results of this experiment.
\rwri shows higher Spearman's $\rho$ and $precision@1, 2, 3$ measures for all implicit rating datasets, than \mfi does.

\begin{table*}[t!]
	\centering
	\caption{
			Comparison of matrix factorization and random walk with restart in using \textbf{implicit feedback} datasets.
			\rwri performs better than \mfi in global ranking and top-1,2,3 prediction, as \rwri presents higher Spearman's $\rho$ and $precision@1, 2, 3$ than \mfi does.
	}
	\label{table:implicit}
	\renewcommand{\tabcolsep}{2.5mm}
	\begin{tabular}{c | c | cc | cc | cc | cc}
		\toprule
		& Performance & \multicolumn{2}{c | }{Spearman's $\rho$} & \multicolumn{2}{c | }{$precision@1$}  & \multicolumn{2}{c | }{$precision@2$}  & \multicolumn{2}{c}{$precision@3$}  \\ \cline{3-10}
		& summary     & Lastfm  & Audioscrobbler  & Lastfm  & Audioscrobbler  & Lastfm  & Audioscrobbler  & Lastfm  & Audioscrobbler   \\
		\midrule
		\mfi  & Lower      & 0.428      & 0.115      & 0.131    & 0.106     & 0.277  & 0.169    & 0.417   & 0.216 \\
		\rwri & Higher       & \textbf{0.558}     & \textbf{0.370}    & \textbf{0.272}  & \textbf{0.149}  & \textbf{0.402}    & \textbf{0.220}        & \textbf{0.517}     & \textbf{0.272} \\
		\bottomrule
	\end{tabular}
\end{table*}

\subsection{Performance of Methods with Global Bias Terms}
\label{sec:exp:bias}
With regard to global bias terms, we present performance measures of matrix factorization and random walk with restart according to the following interests:
(1) whether the bias terms enhance performance, and (2) which method is better between \bmf and \rwrb.
We observe that (1) the bias terms improve recommendation performance, and (2) \bmf performs better than \rwrb.
The experiments are conducted with only explicit feedback rating because it is inappropriate to apply global bias terms in implicit feedback ratings.
The reason is that bias terms and vectors of entities are optimized to be invalid value or even 0 in \bmf with implicit ratings.
This is because the global average rating $\mu$ always offsets the implicit rating values, and thus there are no observed values left for the biases and vectors to learn from.
\bmf approximates $\mu + b_u + b_i + \vect{x}_u^T \vect{y}_i$ to a binarized implicit rating $p_{ui}$; however, $\mu$ and $p_{ui}$ are 1, causing $b_u + b_i + \vect{x}_u^T \vect{y}_i$ to be 0.
In this case, there are no valid observed values.

Bias terms in matrix factorization are known to improve accuracy performance, decreasing RMSE \cite{koren2009matrix}.
We observe that bias terms also enhance global ranking and top-$k$ prediction performance in our experiments.
Tables \ref{table:biasmf} and \ref{table:biasrwr} present the results of our experiments.
Bias terms enhance the performance in both matrix factorization and random walk with restart, as the methods with bias terms show higher Spearman's $\rho$ and $precision@1, 2, 3$ compared to the ones without biases.

\begin{table*}[t]
	\centering
	\caption{
			Comparison of \mfe and \bmf to show the effects of \textbf{global bias terms} in matrix factorization.
			Global bias terms enhance ranking and top-$k$ prediction performance.
			\bmf shows better performance than \mfe in global ranking and top-1, 2, 3 prediction measure, as \bmf presents higher Spearman's $\rho$ and $precision@1, 2, 3$ than \mfe does.
	}
	\label{table:biasmf}
	\renewcommand{\tabcolsep}{0.7mm}
	\begin{tabular}{c | c | ccc | ccc | ccc | ccc}
		\toprule
			& Performance & \multicolumn{3}{c | }{Spearman's $\rho$} & \multicolumn{3}{c | }{$precision@1$}  & \multicolumn{3}{c | }{$precision@2$}  & \multicolumn{3}{c}{$precision@3$}  \\ \cline{3-14}
			& summary     & Movielens  & FilmTrust  & Epinions & Movielens & FilmTrust & Epinions & Movielens & FilmTrust & Epinions & Movielens & FilmTrust & Epinions \\
		\midrule
			\mfe  & Lower      & 0.328      & 0.377      & 0.762    & 0.144     & 0.381     & 0.597    & 0.260     & 0.514     & 0.653    & 0.365     & 0.573     & 0.624    \\
			\bmf & Higher       & \textbf{0.351}      & \textbf{0.424}      & 0.762    & \textbf{0.152}     & \textbf{0.391}     & \textbf{0.624}    & \textbf{0.275}     & \textbf{0.551}     & \textbf{0.665}   & \textbf{0.375}     & \textbf{0.645}     & \textbf{0.631}   \\
		\midrule
			\multicolumn{2}{c|}{Improvement through bias} & 7.0\% & 12.5\% & 0\% & 6.6\% & 2.6\% & 4.5\% & 5.8\% & 7.2\% & 1.8\% &2.7\% & 12.6\% & 1.1\% \\
		\bottomrule
	\end{tabular}
\end{table*}

\begin{table*}[t]
	\centering
	\caption{
			Comparison of \rwr and \rwrb to show the effects of \textbf{global bias terms} in random walk with restart.
			The bias terms enhance ranking and top-$k$ prediction performance.
			\rwrb performs better than \rwr in global ranking and top-1,2,3 prediction measure, as \rwrb presents higher or equal Spearman's $\rho$ and $precision@1, 2, 3$ than \rwr does.
	}
	\label{table:biasrwr}
	\renewcommand{\tabcolsep}{0.7mm}
	\begin{tabular}{c | c | ccc | ccc | ccc | ccc}
		\toprule
			& Performance & \multicolumn{3}{c | }{Spearman's $\rho$} & \multicolumn{3}{c | }{$precision@1$}  & \multicolumn{3}{c | }{$precision@2$}  & \multicolumn{3}{c}{$precision@3$}  \\ \cline{3-14}
			& summary     & Movielens  & FilmTrust  & Epinions & Movielens & FilmTrust & Epinions & Movielens & FilmTrust & Epinions & Movielens & FilmTrust & Epinions \\
		\midrule
			\rwr & Lower       & 0.238      & 0.359      & 0.616    & 0.103     & 0.319     & 0.498    & 0.197     & 0.470     & 0.585    & 0.306     & 0.549     & 0.587   \\
			\rwrb & Higher       & \textbf{0.241}      & \textbf{0.369}      & \textbf{0.625}    & \textbf{0.105}     & \textbf{0.350}     & 0.498    & \textbf{0.199}     & \textbf{0.487}     & 0.585    & \textbf{0.312}     & \textbf{0.552}     & \textbf{0.589}   \\
		\midrule
			\multicolumn{2}{c|}{Improvement through bias} & 1.2\% & 2.8\% & 1.5\% & 1.9\% & 9.7\% & 0\% & 1.0\% & 3.6\% & 0\% &2.0\% & 0.5\% & 0.3\% \\
		\bottomrule
	\end{tabular}
\end{table*}

Our experiment results also show that \bmf performs better than \rwrb, which shows similar tendencies in experiment with explicit rating data in Section \ref{sec:exp:exp}.

\subsection{Performance of Methods with Side Information}
\label{sec:exp:side}
We focus on three aspects of matrix factorization and random walk with restart in using side information:
(1) whether additional information improves recommendation performance,
(2) which method performs better between \coldmf and \coldrwr with side information,
and (3) which method solves the cold start problem better.
In our experiments, the results are observed as follows:
(1) additional information improves or shows modest decrease in the performance with explicit rating data, while the performance decreases with implicit rating data,
(2)  \coldmf performs better with explicit rating data and \coldrwr performs better with implicit ones,
and (3) \coldmf solves the cold start problem better with explicit rating data and \coldrwr is better with implicit ones.

The ranking and top-$k$ prediction performance are improved with auxiliary information and explicit feedback ratings in general.	
Table \ref{table:explicitside} shows the results of our experiments that compare \mfe and \coldmf.
Table \ref{table:explicitsiderwr} shows the performance comparison of \rwr and \coldrwr.
These results present that methods using side information such as \coldmf and \coldrwr perform better in general than methods not using additional information such as \mfe and \rwr.
However, the performance is worse with side information when implicit rating data are given.
Table \ref{table:implicitside} shows that \coldmf performs worse than \mfi.
Table \ref{table:implicitsiderwr} presents that \coldrwr performs worse than \rwri.
These results indicate that additional information needs to be carefully handled for better accuracy.

We observe that \coldmf is better with explicit ratings and \coldrwr performs better with implicit ones.
The result shows the similar tendencies in experiments with explicit or implicit data without side information in Sections \ref{sec:exp:exp} and \ref{sec:exp:imp}.

We compare matrix factorization and random walk with restart in their ability to address the cold start problem, especially to recommend items to new users.		We observe that matrix factorization performs better in solving the cold start problem when explicit ratings are given, while random walk with restart is better with implicit feedback data.
We take the following steps for this experiment.
First, we sample a set of cold start users $\set{U_{cold}}$.
We randomly sample 20\% of total users for the cold users, where the chosen ones have rated more than one item and have side information.
Second, we split training and test set.
The training set contains side information of all users and observed ratings of users who do not belong to $\set{U_{cold}}$.
The test set includes observed ratings of users $\in \set{U_{cold}}$.
Lastly, we measure Spearman's $\rho$ for global ranking performance and $precision@k$ for top-$k$ prediction performance.
Tables \ref{table:coldexplicit} and \ref{table:coldimplicit} show the result that \coldmf better solves the cold start problem with explicit feedback data and \coldrwr is better with implicit ratings.

\begin{table*}[t]
	\centering
	\caption{
			Comparison of \mfe and \coldmf to show whether \textbf{side information} improves the recommendation performance in matrix factorization with restart when \textbf{explicit feedback} data are given.
			The global ranking performance is enhanced with side information and top-$k$ prediction performance also improves except in several cases, as \coldmf presents higher Spearman's $\rho$ in all cases and higher $precision@1, 2, 3$ in almost all cases than \mfe does.
	}
	\label{table:explicitside}
	\renewcommand{\tabcolsep}{0.5mm}
	\begin{tabular}{c | c | ccc | ccc | ccc | ccc}
		\toprule
			& Performance & \multicolumn{3}{c | }{Spearman's $\rho$} & \multicolumn{3}{c | }{$precision@1$}  & \multicolumn{3}{c | }{$precision@2$}  & \multicolumn{3}{c}{$precision@3$}  \\ \cline{3-14}
			& summary     & Movielens  & FilmTrust  & Epinions & Movielens & FilmTrust & Epinions & Movielens & FilmTrust & Epinions & Movielens & FilmTrust & Epinions \\
		\midrule
			\mfe  & Lower      & 0.328      & 0.377      & 0.762    & 0.144     & 0.381     & 0.597    & 0.260     & 0.514     & \textbf{0.653}    & 0.365     & 0.573     & \textbf{0.624}    \\
			\coldmf & Higher       & \textbf{0.361}      & \textbf{0.416}      & \textbf{0.793}    & \textbf{0.147}     & \textbf{0.403}     & \textbf{0.650}    & \textbf{0.269}     & \textbf{0.524}     & 0.628    & \textbf{0.375}     & \textbf{0.574}     & 0.577   \\
		\midrule
			\multicolumn{2}{c|}{Improvement through side info} & 10.0\% & 10.3\% & 4.0\% & 2.0\% & 5.8\% & 8.9\% & 3.5\% & 1.9\% & -3.8\% &2.7\% & 0.2\% & -7.6\% \\
		\bottomrule
	\end{tabular}
\end{table*}

\begin{table*}[t]
	\centering
	\caption{
			Comparison of \rwr and \coldrwr to show whether \textbf{side information} improves the recommendation performance in random walk with restart when \textbf{explicit feedback} data are given.
			The global ranking performance is enhanced with side information and top-$k$ prediction performance also improves except in several cases.
			\coldrwr presents higher Spearman's $\rho$ in all cases and higher $precision@1, 2, 3$ in almost all cases than \rwr does.
	}
	\label{table:explicitsiderwr}
	\renewcommand{\tabcolsep}{0.5mm}
	\begin{tabular}{c | c | ccc | ccc | ccc | ccc}
		\toprule
		& Performance & \multicolumn{3}{c | }{Spearman's $\rho$} & \multicolumn{3}{c | }{$precision@1$}  & \multicolumn{3}{c | }{$precision@2$}  & \multicolumn{3}{c}{$precision@3$}  \\ \cline{3-14}
		& summary     & Movielens  & FilmTrust  & Epinions & Movielens & FilmTrust & Epinions & Movielens & FilmTrust & Epinions & Movielens & FilmTrust & Epinions \\
		\midrule
		\rwr  & Lower      & 0.238      & 0.359      & 0.616  & 0.103    & 0.319     & 0.498     & 0.197    & \textbf{0.470}     & \textbf{0.585}     & 0.306    & \textbf{0.549}     & \textbf{0.587}      \\
		\coldrwr & Higher       & \textbf{0.239}      & \textbf{0.384}      & \textbf{0.678}    & 0.103     & \textbf{0.357}     & \textbf{0.578}    & 0.197     & 0.438     & 0.573    & 0.306     & 0.540     & 0.542   \\
		\midrule
		\multicolumn{2}{c|}{Improvement through side info} & 0.4\% & 7.0\% & 10.0\% & 0\% & 11.9\% & 8.9\% & 0\% & -6.9\% & -2.1\% &0\% & -6.2\% & -7.7\% \\
		\bottomrule
	\end{tabular}
\end{table*}

\begin{table*}[t]
	\centering
	\caption{
			Comparison of \mfi and \coldmf to show whether \textbf{side information} improves the recommendation performance with \textbf{implicit rating}.
			The performance decreases with side information.
			\coldmf shows worse performance than \mfi in global ranking and top-1, 2, 3 prediction measure, as \coldmf presents lower Spearman's $\rho$ and $precision@1, 2, 3$ than \mfi does.
	}
	\label{table:implicitside}
	\renewcommand{\tabcolsep}{7.7mm}
	\begin{tabular}{c | c | c | c | c | c}
		\toprule
			& Performance & Spearman's $\rho$ & $precision@1$ & $precision@2$ & $precision@3$ \\		 \cline{3-6}
			& summary     & Lastfm         & Lastfm      & Lastfm      & Lastfm      \\
					\midrule
			\mfi  & Higher      & \textbf{0.428}          & \textbf{0.131}       & \textbf{0.277}       & \textbf{0.417}       \\
			\coldmf & Lower       & 0.388          & 0.130       & 0.263       & 0.397    \\
		\midrule
			\multicolumn{2}{c|}{Improvement through side info} & -9.4\% & -0.8\% & -5.1\% & -4.8\% \\
		\bottomrule
	\end{tabular}
\end{table*}

\begin{table*}[t]
	\centering
	\caption{
			Comparison of \rwri and \coldrwr to show whether \textbf{side information} improves or decrease the recommendation performance when \textbf{implicit rating} data are given.
			The performance decreases with side information.
			\coldrwr shows less performance than \rwri in global ranking and top-1,2,3 prediction measure, as \coldrwr presents lower Spearman's $\rho$ and $precision@1, 2, 3$ than \rwri does.
	}
	\label{table:implicitsiderwr}
	\renewcommand{\tabcolsep}{7.7mm}
	\begin{tabular}{c | c | c | c | c | c}
		\toprule
		& Performance & Spearman's $\rho$ & $precision@1$ & $precision@2$ & $precision@3$ \\		 \cline{3-6}
		& summary     & Lastfm         & Lastfm      & Lastfm      & Lastfm      \\
		\midrule
		\rwri  & Higher      & \textbf{0.558}          & \textbf{0.272}       & \textbf{0.402}       &\textbf{0.517}       \\
		\coldrwr & Lower       & 0.544          & 0.242       & 0.385       & 0.505    \\
		\midrule
		\multicolumn{2}{c|}{Improvement through side info} & -2.6\% & -1.1\% & -4.3\% & -2.4\% \\
		\bottomrule
	\end{tabular}
\end{table*}

\begin{table*}[t]
	\centering
	\caption{
			Comparison of \coldmf and \coldrwr to evaluate which method better solves the \textbf{cold start} problem when \textbf{explicit ratings} are given.
			\coldmf performs better in this case.
			\coldmf shows better performance than \coldrwr in global ranking and top-1,2,3 prediction measure, as \coldmf presents higher Spearman's $\rho$ and $precision@1, 2, 3$ than \coldrwr does.
	}
	\label{table:coldexplicit}
	\renewcommand{\tabcolsep}{0.9mm}
	\begin{tabular}{c | c | ccc | ccc | ccc | ccc}
		\toprule
		& Performance & \multicolumn{3}{c | }{Spearman's $\rho$} & \multicolumn{3}{c | }{$precision@1$}  & \multicolumn{3}{c | }{$precision@2$}  & \multicolumn{3}{c}{$precision@3$}  \\ \cline{3-14}
		& summary     & Movielens  & FilmTrust  & Epinions & Movielens & FilmTrust & Epinions & Movielens & FilmTrust & Epinions & Movielens & FilmTrust & Epinions \\
		\midrule
		\coldmf  & Higher      & \textbf{0.310}      & \textbf{0.248}      & \textbf{0.640}  & \textbf{0.050}    & \textbf{0.162}     & \textbf{0.369}     &\textbf{0.092}    & \textbf{0.255}     & \textbf{0.419}     & \textbf{0.115}    & \textbf{0.288}     & \textbf{0.448}      \\
		\coldrwr & Lower       & 0.231      & 0.201      & 0.399    & 0.032     & 0.146     & 0.362    & 0.064     & 0.237     & 0.416    & 0.105     & 0.279     & 0.446   \\
		\bottomrule
	\end{tabular}
\end{table*}

\begin{table*}[t]
	\centering
	\caption{
			Comparison of \coldmf and \coldrwr to evaluate which method better solves the \textbf{cold start} problem when \textbf{implicit ratings} are given.
			\coldrwr performs better in this case.
			\coldrwr shows better performance than \coldmf in global ranking and top-1,2,3 prediction measure, as \coldrwr presents higher Spearman's $\rho$ and $precision@1, 2, 3$ than \coldmf does.
	}
	\label{table:coldimplicit}
	\renewcommand{\tabcolsep}{7.7mm}
	\begin{tabular}{c | c | c | c | c | c}
		\toprule
		& Performance & Spearman's $\rho$ & $precision@1$ & $precision@2$ & $precision@3$ \\		 \cline{3-6}
		& summary     & Lastfm         & Lastfm      & Lastfm      & Lastfm      \\
		\midrule
		\coldmf  & Lower      & 0.207          & 0.026       & 0.041       & 0.058       \\
		\coldrwr & Higher       & \textbf{0.440}          & \textbf{0.070}       & \textbf{0.100}       & \textbf{0.149}    \\
		\bottomrule
	\end{tabular}
\end{table*}

\subsection{Discussion}
\label{sec:dis}
\input{500discussion}

%% file: 500discussion.tex
What are the reasons for good or bad performance of MF and RWR in various settings of recommendations?
We observe interesting tendencies over various performance measures in our experiments:
\begin{itemize}
	\item Matrix factorization performs better with explicit rating datasets, while random walk with restart performs better with implicit ones.
	\item Global bias terms improve recommendation performance.
	\item Additional information enhance performance of recommendations with explicit rating data, while it makes worse performance with implicit rating data.
\end{itemize}

Why matrix factorization conforms well to explicit ratings while random walk with restart applies well to implicit ones?
Random walk with restart is influenced by proximity of nodes or the number of hops, while explicit rating scores occasionally contradict the concept of the proximity.
For example, random walk with restart gives a higher RWR score on a 1-hop negative item (e.g. an item that a user $u$ gives a low rating) than a 3-hop positive item (e.g. an item preferred by another user whose taste is similar to $u$).
Implicit data only include users' positive behavior on items and do not include a shortcut between negatively related nodes.
Therefore, random walk with restart fits well with implicit data rather than explicit ones.
This explains A1, A2, A4, A6, and A7 in the question and answer list at the beginning page in Section \ref{sec:exp}. As regards A4, \bmf performs better than \rwrb because we use only explicit ratings in comparison of them as explained in Section~\ref{sec:exp:bias}.

	Global bias terms enhance the global ranking and top-$k$ prediction performance in both matrix factorization and random walk with restart.
	Bias factors improve the performance because the methods learn not only user-item interactions but also nodes' specific properties accounting for much of the variation in observed ratings \cite{koren2009matrix}.
	This explains A3 in the question and answer list.

	Additional information is used well with explicit feedback data, but not with implicit ones.
	Our conjecture is that side information is noisy, and the noise further debases the quality of information in implicit feedback ratings that are also noisy.
	This possibly explains A5 in the question and answer list.

%% file: 700conclusion.tex
We provide a comparative study of matrix factorization (MF) and random walk with restart (RWR) in recommender system.
We suggest four tasks according to various recommendation scenarios.
We suggest recommendation methods based on MF and RWR,
and show that each scenario has its corresponding MF and RWR methods.
Especially, we devise a new RWR method using global bias terms, and the new method improves the recommendation performance.
We provide extensive experimental results that compare MF and RWR for the various tasks and explain insight for the reasons behind the results.
We observe that MF and RWR behave differently according to whether input ratings are explicit or implicit.
We also observe that global popularities of nodes represented as biases are useful in improving the recommendation performance. 
Finally, we observe that side information improves recommendation quality with explicit feedback, while degrades it with implicit feedback.